\documentclass[manuscript]{aastex631}
\usepackage{CJK}
\usepackage{amsmath}
\shorttitle{Mini Coronal Dimmings}
\shortauthors{Wang et al.}
\graphicspath{{./}{figures/}}

\begin{document}
\begin{CJK*}{UTF8}{gbsn}
\title{Observations of Mini Coronal Dimmings Caused by Small-scale Eruptions in the Quiet Sun}

\correspondingauthor{Rui Wang}
\email{rwang@swl.ac.cn; liuxying@swl.ac.cn}

\author[0000-0001-5205-1713]{Rui Wang (王瑞)}
\affiliation{State Key Laboratory of Space Weather, National Space Science Center, Chinese Academy of Sciences, Beijing 100190, People's Republic of China}

\author[0000-0002-3483-5909]{Ying D. Liu}
\affiliation{State Key Laboratory of Space Weather, National Space Science Center, Chinese Academy of Sciences, Beijing 100190, People's Republic of China}
\affiliation{University of Chinese Academy of Sciences, Beijing 100049, People's Republic of China}

\author[0000-0002-4016-5710]{Xiaowei Zhao}
\affiliation{Key Laboratory of Space Weather, National Satellite Meteorological Center (National Center for Space Weather), China Meteorological Administration, Beijing 100081, People's Republic of China}
\affiliation{School of Earth and Space Sciences, Peking University, Beijing 100871, People's Republic of China}

\author[0000-0001-8188-9013]{Huidong Hu}
\affiliation{State Key Laboratory of Space Weather, National Space Science Center, Chinese Academy of Sciences, Beijing 100190, People's Republic of China}



\begin{abstract}

Small-scale eruptions could play an important role in coronal heating, generation of solar energetic particles (SEPs), and mass source of the solar wind. However, they are poorly observed, and their characteristics, distributions, and origins remain unclear. Here a mini coronal dimming was captured by the recently launched Solar Orbiter spacecraft. The observations indicate that a minifilament eruption results in the dimming and takes away approximately $(1.65\pm0.54)\times10^{13}$ g of mass, which also exhibits similar features as the sources of SEP events. The released magnetic free energy is of the order of $\sim10^{27}$ erg. Our results suggest that weak constraining force makes the flux rope associated with the minifilament easily enter a torus-unstable domain. We discuss that weak magnetic constraints from low-altitude background fields may be a general condition for the quiet-Sun eruptions, which provide a possible mechanism for the transport of coronal material and energy from the lower to the middle or even higher corona.

\end{abstract}



\section{Introduction} \label{sec:intro}

Solar eruptive activities, including flares, eruptive filaments, coronal mass ejections (CMEs), and coronal jets, are widely distributed in energy scales. The observed solar eruptions span a range of at least a factor of $10^8$ in energy \citep{2012Schrijver}. Large, highly energetic ``X-class'' eruptions can release energy substantially exceeding $\sim10^{33}$ erg. With the advent of high-resolution observations, the scaled-down versions of the eruptive activities are getting more attention, e.g., nanoflares \citep{2021Bahauddin}, minifilaments \citep{2000Wangjx,2009Innes,2010Innes,2015Sterling,2016Panesar,2018Panesar,2019McGlasson}, and the newly discovered ``campfires'' \citep{2021Berghmans}, which are at lower energies probably below the current detection limit. Even-smaller-scale eruptive structures have been speculated to exist \citep{1986Hermans, 2016Sterling}. The latest observations of Parker Solar Probe near the Sun \citep{2019McComas} indicate that small solar energetic particle (SEP) events, which cannot be captured by 1 au spacecraft but are only observable close to the Sun, may be much more common than previously thought. Where do these small SEP events come from? Is it possible that they are generated by these small-scale solar eruptions?

\citet{1993Jackson} and \citet{1994Webb} suggest that the ratio of the annualized CME to solar wind mass flux is no more than $16\%$, and \citet{2017Lamy} indicate that this ratio is only up to $6\%$. However, it should be noted that the mass of these counted CMEs is usually in the range of $\sim10^{14}-10^{15}$ g. Then what about small solar eruptions with less mass? \citet{2000Aschwanden} and \citet{2012Schrijver} indicate that the events at the lowest energies of $\sim10^{24}$ erg can occur millions of times each day. Namely, smaller eruptions generally have higher frequency of occurrence. How do the small-scale eruptions, with the advantages of large numbers and high occurrence rate, contribute to the mass of the solar wind? Most of our knowledge of the energy-mass relation of eruptions is based on the observations of large-scale eruptions. It remains unclear what the actual relationship is between the mass and energy of small-scale eruptions in the source region.

 Coronal dimmings are considered to be an important characteristic of solar eruptions, which can provide mass information of eruptions \citep{1976Rust,1998Thompson}. It is generally accepted that they are caused by plasma evacuation during the eruption of a CME \citep{2000Harrison}. On the Sun, the dimming regions correspond to the footprint of the CME \citep{2022Jinmeng}. Upflowing expanding plasma has been observed in the dimming region \citep{2001Harra,2012Tianh}. \citet{2000Harrison} find that the mass loss in the coronal dimmings is on the same order as the estimated mass of the associated CMEs. Therefore, the mass of small-scale eruptions can also be estimated by their associated coronal dimmings. However, small-scale coronal dimmings have been poorly studied. A small dimming associated with an eruption is not easy to identify, since the eruption itself and the associated characteristic structures, such as post-flare arcades, flare ribbons, and coronal waves, are difficult to recognize under the previous observational limits. Therefore, it is difficult to determine whether the observed dimming is related to an eruption or not. In short, we do not know much about how small-scale coronal dimmings are produced, what their characteristics are, and how they are distributed.

In this Letter, we identify the characteristic structures associated with a small-scale eruption with greater certainty, owing to the 17.4 nm EUV High Resolution Imager (HRI$_{EUV}$) of the Extreme Ultraviolet Imager \citep[EUI;][]{2020Rochus} \footnote{We used level 2 (L2) EUI data, which can be accessed via \url{http://sidc.be/EUI/data/releases/202112\_release\_4.0}. Information about the data processing can be found in the release notes. DOI: \url{https://doi.org/10.24414/s5da-7e78}.} on board Solar Orbiter \citep{2020Muller}. We estimate the evacuated mass and the released energy of the eruption. A survey of mini coronal dimmings from Solar Orbiter observations is also carried out for the first time to investigate the distributions and occurrence frequency.
These results provide crucial information for understanding small-scale eruptions and their contributions to SEPs, the solar wind, and space weather.

\section{Results}\label{sec2}

\subsection{High-resolution Observations by Solar Orbiter}

On 2020 May 20 21:20 universal time (UT), a small-scale eruption with coronal dimming occurred near the central meridian, which was captured by the recently launched Solar Orbiter spacecraft. At this moment, Solar Orbiter was located at a distance of 0.612 au from the Sun during its perihelion pass with a separation angle of $\sim17^\circ$ from the Sun-Earth line (Figure~1a) and was almost in quadrature with STEREO-A. From this vantage point, the HRI$_{EUV}$ angular pixel size of $0^{\prime\prime}.492$ corresponds to 217 km on the solar surface. Even under such a high spatial resolution, the eruption itself is still too small (length scale $\sim$10 Mm) to be noticed. By contrast, the associated coronal dimming is relatively easier to be observed (length scale $\sim$34 Mm; see Figure~1b, c, d, and the animation of Figure 1; images for making the video aligned by means of a cross-correlation method to remove the effect of jitter in the data). Solar Dynamics Observatory \citep[SDO;][]{2012Pesnell} allows us to carry out a joint observation of this small event. The HRI$_{EUV}$ of EUI provides a more distinct EUV image than the Atmospheric Imaging Assembly \citep[AIA;][]{2012Lemen} on board SDO. Figure~1e shows clearly discernible strip structures in an HRI$_{EUV}$ bandpass centered at 17.4 nm. The distinct bright strips traced by the red dotted curves are considered as the post-eruption arcades, which are generally associated with newly reconnected field lines caused by the rise of filaments during the eruption process \citep{1995Shibata,2011Reeves}. We remap the SDO's AIA image at 1600 \AA~and the Helioseismic and Magnetic Imager \citep[HMI;][]{2012Scherrer,2012Schou} line-of-sight (LOS) magnetic fields from the SDO view to the Solar Orbiter view. It shows that the footpoints of the bright strips are also brightened in AIA 1600 \AA~(Figure~1f), which are considered as the flare ribbons following the eruption. The ribbons are aligned with the relatively strong photospheric magnetic fields where the post-eruption arcades are rooted (Figure~1g).

\begin{figure}[h]
   \centering
   \includegraphics[width=1.0\textwidth]{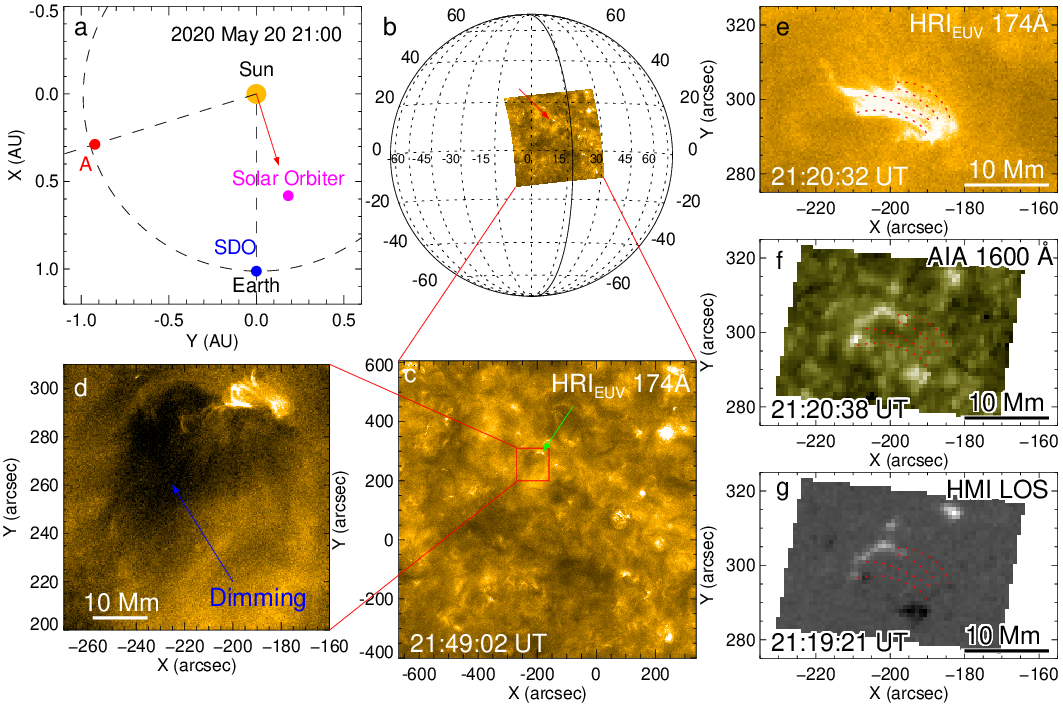}
   \caption{Joint observations of Solar Orbiter and SDO on 2020 May 20. (a) Heliospheric positions of planets and spacecraft. (b) Location of the mini coronal dimming (red arrow) in HRI$_{EUV}$ 174 \AA. (c) and (d) Enlarged views of the dimming at different scales. (e) Post-eruption arcades (traced by red dotted curves) in HRI$_{EUV}$ 174 \AA, with location shown by the green arrow in (c). (f) Flare ribbons in SDO/AIA 1600 \AA. (g) SDO/HMI LOS magnetic fields. An animation of this figure is available. It shows the EUI animation of the dimming from 21:20 to 22:16 UT without annotations. Its real-time duration is 5 s. It has an FOV slightly larger than (d) but smaller than (c).}\label{fig1}
\end{figure}

\subsection{Temperature and Mass in the Mini Dimming Region}

We use a differential emission measure (DEM) analysis to examine the temperature and density properties of the coronal dimming. In this work, we adopt the ``simple\_reg\_dem.pro'' routine \citep{2020Plowman} in SSW packages to compute the DEM with the AIA EUV images at six passbands (94, 131, 171, 193, 211, and 335~\AA) in the temperature range of 5.5 $\leq$ log$_{10}(T)$ $\leq$ 7.5. This algorithm is fast and relatively free from idiosyncrasies, which is suitable for long time series analysis of the DEM of numerous pixels.

Figure~2 shows the evolution of EM for 0.7 MK plasma. We select a heart-shaped area (contoured in red), which almost contains all the dimming areas during the eruption and meanwhile excludes the bright plasma (i.e., the core region of the eruption under the post-eruption arcades within the black contours). The cool-color region within the heart-shaped area (lower density) gradually expanded (Figures~2a-c). At 21:07 UT, the beginning of the eruption, there are some preexisting dim locations in the corona (within the heart-shaped area) that are not associated with the eruption. At 21:40 UT, we can observe a small dimming area appearing near the core region. Then, the dimming area rapidly expanded around 22:00 UT and filled the whole heart-shaped area by 22:26 UT.

Figure 2e shows that the changes of the normalized intensities of the flare region exhibit a trend of increase before 21:00 UT. Figure 2f shows a significant decrease in the intensity curve of the heart-shaped area after 21:40 UT which is attributed to the occurrence of coronal dimming in that area. Prior to the flares associated with the dimming, the flare region exhibited continuous EUV brightenings, which may have been caused by magnetic reconnection resulting from the motion of the photospheric magnetic field.

Figure 2g shows the averaged EM within the dimming region as a function of time, which is calculated using
\begin{equation}\label{EQ1}
    EM = \int DEM(T)dT.
\end{equation}
The EM has an obvious decrease after 21:40 UT and almost keeps constant after 22:00 UT, which indicates that coronal plasma probably escaped from the dimming region during the eruption. The EM-weighted median temperature (red curve) in Figure~2e is defined as
\begin{equation}\label{tem}
    \overline{T} = \frac{\int DEM(T)\times TdT}{\int DEM(T)dT}.
\end{equation}
It almost keeps constant below $10^{6.5}$ K, which is a typical temperature in a dimming region \citep{2012Chengxin}. This implies that the EUV dimming mainly resulted from plasma depletion during the eruption. By the DEM analysis, the mass loss within the heart-shaped area is estimated to be $(1.65\pm0.54)\times10^{13}$ g (see Appendix). This mass is 1 to 2 orders lower than the normal CME mass of $\sim10^{14}-10^{15}$ g \citep{2010Vourlidas}.

\begin{figure}[h]
    \centering
    \includegraphics[width=1.0\textwidth]{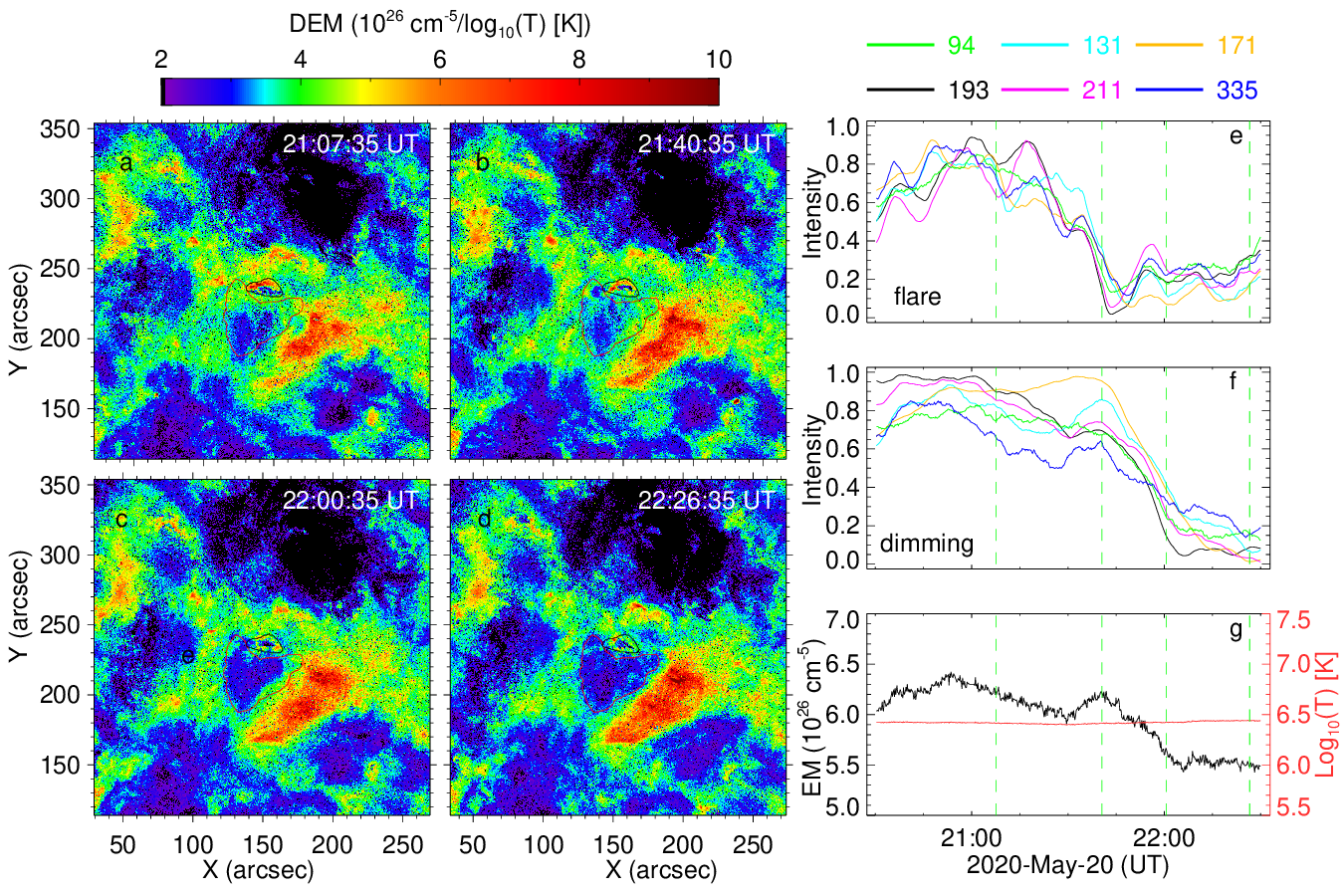}
    \caption{DEM analysis of the coronal dimming. (a-d) Time evolution of AIA 0.7 MK DEMs at four typical times. (e-f) EUV light curves of the normalized integrated intensities of the flare region (black contour) and the dimming region within the heart-shaped area (red contour). (g) Averaged EM (black) and EM-weighted median temperature (red) within the dimming region as a function of time. The green vertical dashed lines mark the corresponding times in (a-d).}\label{fig2}
\end{figure}

\subsection{Minifilament Eruption and the Associated Magnetic Fields}

We do not observe a filament structure in EUI/HRI$_{EUV}$ 174~\AA. Fortunately, the 304~\AA~data from SDO/AIA can provide supplementary information for observing filament structures that may not be visible in EUI/HRI$_{EUV}$ 174~\AA~images. Although AIA has a relatively lower spatial resolution, we were able to track the erupting filament structure at 304 \AA~as it rose above the bright structures at the source region of the eruption (see the animation of Figure 3a). The S-shaped filament (contoured by the dotted line) is probably related to a rising, twisted flux-rope structure (Figure~3a). Figure~3b shows the time-distance profile of the rising filament along the slit (3 pixel width) in Figure~3a. The pattern in Figure~3b implies that the filament rises in a rolling manner expanding to higher altitudes. The maximum of the local projected speed is around 12 km $s^{-1}$; the average projected speed is only around 3.6 km $s^{-1}$. Although it is rather slow compared with regular filament eruptions \citep{2016Rui2,2020Chengxin}, this speed is considered to be relatively typical for minifilament eruptions \citep{2000Wangjx,2020Panesar,2022SterlingB}. \citet{2020Panesar} indicate that the speed of the quiet-region jet-making minifilament eruptions is at least $\sim$ 1 km s$^{-1}$ during the slow-rise phase. The average projected speed in our case is comparable to that. \citet{2022SterlingB} suggest that the slow speed reported by \citet{2020Panesar} may be attributed to a projection effect caused by differences in viewing angles. On the other hand, time-distance plot helps us to roughly determine the start time of the eruption, which is around 21:15 UT.

\begin{figure}[h]
    \centering
    \includegraphics[width=1.0\textwidth]{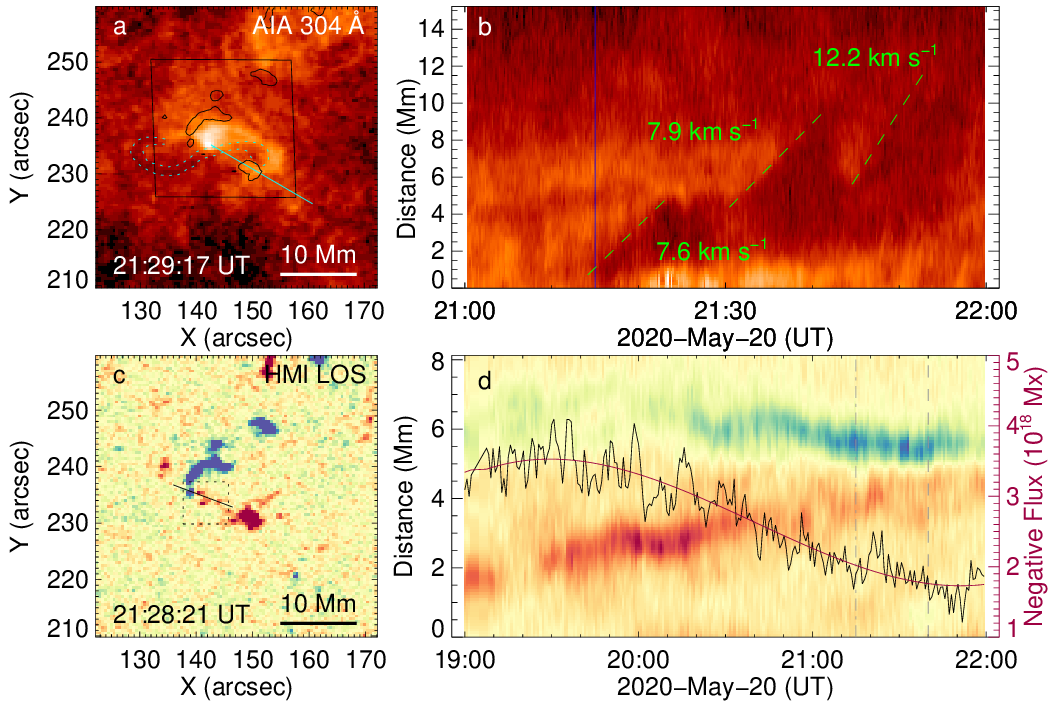}
    \caption{Small-scale filament eruption and the associated magnetic cancellation. (a) Rising minifilament in AIA 304 \AA~(contoured by the cyan dotted line). The black square corresponds to the boundary for extrapolation in Figure~4. (b) Time-distance profile of the rising filament along the slit (cyan) in (a). The blue vertical line marks the onset of the filament rise. (c) Positive (blue) and negative (red) magnetic fields scaled to $\pm150$ G corresponding to the black contours in (a). (d) HMI time-distance map along the slit in (c). The integrated negative magnetic flux as a function of time computed inside the dotted box in (c) is overplotted in black. A least-squares polynomial fit curve (red) indicates the trend of the flux. The vertical dotted-dashed and dashed lines correspond to the times of the flare and dimming, respectively. An animation of this figure is available. It shows the AIA 304 \AA~animation from 21:00 to 21:57 UT, and has an FOV larger than (a). Its real-time duration is 7 s. The green arrow in the first frame of the animation indicates the approximate position of the rising filament.}\label{fig3}
\end{figure}

Figure~3c shows the distribution of the SDO/HMI LOS magnetic fields. The length scale of the major magnetic polarity is around 10 Mm. Filaments with lengths ranging from 10 to 20 Mm are generally classified as minifilaments \citep{2000Wangjx,2015Sterling}. Therefore, the filament in our case should fall into this category. The fast moving negative magnetic polarities cancel the continually emerging positive polarities in the dotted box region of Figure~3c. We make a narrow slit (3 pixel width) along the converging direction of the moving negative polarities. The time-distance plot of Figure~3d displays that the negative magnetic polarities converge with the emerging positive ones around 19:30 UT. This is similar to the magnetic cancellation in a normal active region. Flux emergence and magnetic cancellation are a coupled process and have always been observed to appear together and interact with each other \citep{2018Rui2,2022Rui,2019Chintzoglou}. Due to the emergence of the positive flux, we can only calculate the change of the negative flux in the dotted box region (black curve in Figure~3d). The red curve in Figure~3d presents a significant drop in the flux in the hour before the eruption (the vertical dotted-dashed line). The time-distance plot patterns indicate a closer convergence of the positive and negative magnetic poles at the onset of the eruption. Moreover, the calculation region corresponds to the EUV brightenings at 304 \AA~observed after the onset of the filament eruption (see Figure 3a and 3c). These results suggest that magnetic cancellation occurred before the eruption and likely resulted in the eruption. Magnetic cancellation is generally related to the buildup of a magnetic flux rope and is also related to magnetic reconnection resulting in the final eruption \citep{1989Vanballegooijen,2001Moore,2018Rui2,2022Rui}. In fact, eruptions caused by cancellation in the quiet Sun often manifest as minifilament eruptions with coronal jets \citep{2015Sterling,2016Panesar,2018Panesar,2019McGlasson,2021Muglach}.

Figure~4a shows that the extrapolated magnetic fields exhibiting twisted morphology along the polarity inversion line (PIL), which is regarded as a flux rope. A magnetic dip of the flux rope is just above the cancellation region (see Figure~4a). The cool, dense filament material is probably located here via a magnetic tension force \citep{2000Gilbert}. Meanwhile, it is often related to the appearance of bald patches in the PIL. The field lines threading the bald patch form a separatrix surface where reconnection preferentially occurs \citep{1993Titov}. Figure 4a shows a hook structure of the flux rope in the southeast corresponding to the EUV filament at 304 \AA~in Figure~4b, which should be part of the entire filament and is located at a higher altitude above the bright magnetic reconnection region.

We adopt two different methods to reconstruct the flux-rope structure, i.e., a magnetohydrostatic (MHS) method \citep{2018Zhuxs} and a widely used optimization approach \citep{2004Wiegelmann}. Both methods give similar twisted S-shaped flux-rope structures. Figure~4a only displays the reconstructed results of the MHS for its better performance in the lower corona.

SDO/HMI does not provide the direct boundary data for extrapolation in our case. We adopt the ``bvec2cea.pro'' routine in SSW packages to convert the disambiguated full-disk vector magnetic field ``hmi.B\_720s'' series from the native CCD coordinates to the cylindrical equal area heliographic coordinates \citep{2014Hoeksema}, which is appropriate for extrapolation. We centered the remapped data on Carrington coordinates $(22^\circ W,12^\circ N)$ where the major magnetic polarities are located. The ``bvec2cea.pro'' routine uses a radial-acute method \citep{2014Hoeksema} to resolve the 180$^\circ$ azimuthal uncertainty in the transverse field direction. Meanwhile, we cut out a relatively small region from the remapped data for extrapolation in order to keep flux balance of the boundary condition (see Figure~3a).

The magnetic free energy is also calculated by the two methods, which give relatively similar results, i.e., $\Delta E_{free}^{MHS} = (4.87\pm0.80)\times10^{27}$ erg, and $\Delta E_{free}^{NLF} = (3.10\pm0.68)\times10^{27}$ erg. An eruption of $\Delta E_{free}\sim10^{27}$ erg is classified as a microflare by the flare-energy power law summarized from early flare statistics \citep{2000Aschwanden,2012Schrijver}.

\begin{figure}[h]
    \centering
    \includegraphics[width=0.8\textwidth]{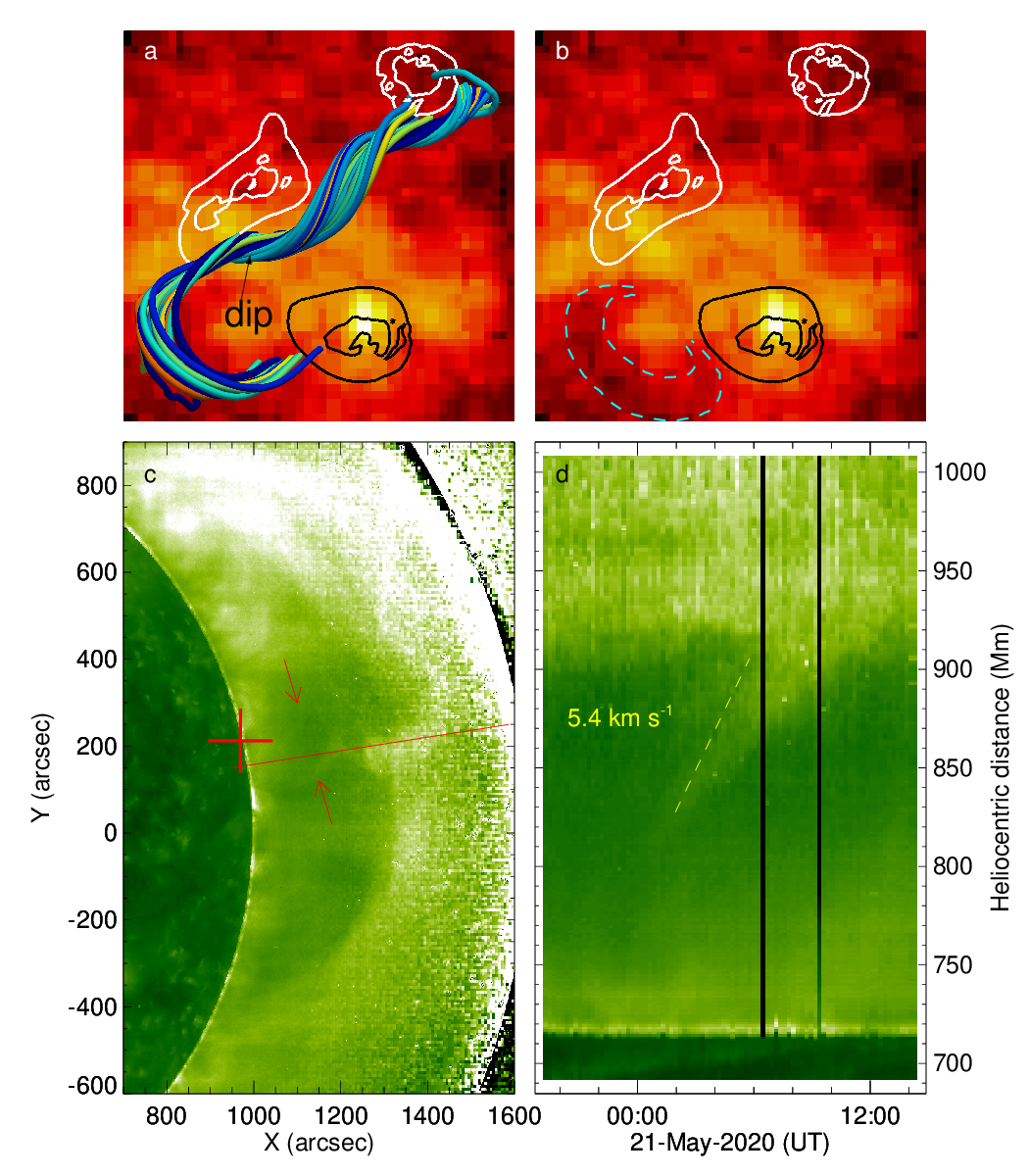}
    \caption{Magnetic topological structure of the small-scale filament (prominence) and its ascending process in the STEREO-A/EUVI view. (a) Reconstructed flux-rope structure with the EUV background at 304 \AA. The positive (white) in [20, 50] G and negative (black) fluxes in [-20, -50] G are overplotted. The rising part of the minifilament is contoured in cyan dashed lines in (b). (c) Side view from STEREO-A/EUVI of the ascending filament (red arrows) along a red radial line, which shows the slit for the time-distance map in (d). The red cross marks the approximate position of the eruption source. An animation of this figure is available. The EUVI animation starts at 2020 May 20 18:06 UT and ends at 2020 May 22 01:58 UT. Its real-time duration is 7 s. With the exception of the arrows, it has the same annotations as (c) but a smaller FOV. The red cross disappears as the source rotates to the far side of the Sun after 2020 May 21 08:00 UT.}\label{fig4}
\end{figure}

\section{Summary and Discussion}
Our results indicate that the small-scale coronal dimming results from an eruption of a minifilament, which is likely to have similar eruption features (e.g., post-eruption arcades and flare ribbons), trigger mechanisms (associated with magnetic cancellation), and magnetic topology (e.g., flux-rope structures and magnetic dips) to large eruptions. We compare the mass and energy loss with previous statistics on the power-law relationship between the flare X-ray fluence and CME mass \citep{2013Drake}. Our results fit the energy-mass relation, i.e., an eruption of $\Delta E_{free}\sim10^{27}$ erg carries the mass in the order of $\sim10^{13}$ g.

\citet{2015Sterling} and \citet{2016Sterling} indicate that solar coronal jets probably result from eruptions of small-scale filaments. \citet{2016Panesar,2018Panesar} suggest that the continuous cancellation can destabilize minifilaments and result in eruptions, where internal magnetic reconnection occurs. The minifilament as shown here has similar size and takes comparable energy of $\sim10^{27}$ erg with the small filaments for driving jets. Moreover, the EUI/HRI$_{EUV}$ observations present jet-like structures at the beginning of the eruption and post-flare arcades (see animation of Figure 1). Therefore, this eruption is likely to have similar formation and trigger mechanisms as coronal jets. \citet{2019McComas} indicate that small, longitudinally distributed SEP events are associated with jet-like coronal emissions, some of which are not observed at 1 au but only detected close to the Sun. Thus, the mini eruptions as shown here become the potential candidates for these small SEP events. There have been quite a few reports of small-scale eruptions with coronal dimming \citep[e.g., ][]{2009Innes,2010Innes,2012YangJY,2013Innes,2014ZhangQM,2022SterlingB}, but due to limitations in observation conditions, it is difficult to discern the intricate features of the eruption sources associated with these events, such as the post-flare arcades or the temperature and density of the corona. These difficulties result in challenges in determining, for instance, whether the coronal dimming is a long-lasting effect caused by mass evacuation or just a wave-like dimming and whether they are associated with the aforementioned on-disk jets. The combined observations of Solar Orbiter and SDO in this work have helped to address some of these challenges.

The small-scale eruptions are omnipresent features in the quiet Sun, estimated with as many as 1400 events per day over the whole Sun \citep{2009Innes,2022Madjarska}. Meanwhile, the total number of mini dimmings (associated with eruption or not) for a full disk in a day \citep[e.g.,][]{2012Alipour} is much more than twice of \citet{2009Innes}. We make a survey of mini coronal dimmings from EUI/HRI$_{EUV}$ high-resolution observations as shown in Table 1, which gives the approximate time and location of the dimmings. These dimmings are all associated with small-scale eruptions. Although the valid EUI/HRI$_{EUV}$ observations are only available on certain days in certain months, we still identify at least one dimming event from the valid 1 or 2 hr observation periods for a day. We only select the relatively obvious events in the field of view (FOV) of a local area. There should be more events on a global scale.

On the other hand, we check the torus instability of the flux rope in the lower coronal (below $\sim$70 Mm) by the decay index $n=-\partial ln \|\mathbf{B}_p\|/\partial ln R$ \citep{2006Kliem}, where $\|\mathbf{B}_p\|$ is an external poroidal field and $R$ is the height of the flux-rope axis's apex above the photosphere, which shows a low critical height $\sim$8 Mm above the PIL. This height is favorable for the escape of a CME from its source region. Understandably, weaker quiet-Sun magnetic fluxes result in weaker constraining force above the flux rope. However, it does not mean that CMEs can successfully escape into interplanetary space. Figure~4c and 4d show that the minifilament propagated slowly along a radial direction in the lower corona with an average speed only $\sim$5.4 km s$^{-1}$. In fact, the slow speed is not unexpected, which is also attributed to the weaker quiet-Sun magnetic fluxes. Namely, they generate a non-strong toroidal current inside the flux rope, which results in a relatively weak upward hoop force. On the other hand, quiet-Sun eruptions in the current solar cycle are mainly under streamer belts where closed background fields are generally dominated (see Figure 4c). Nonetheless, it remains challenging to determine whether quiet-Sun eruptions are confined or eruptive. In our case, we find many narrow blobs moving outwards in the STEREO-A/COR2 view. However, due to the lack of usable COR1 data, we cannot determine the exact origin of these blobs. Regardless, quiet-Sun eruptions are more likely to get into torus-unstable regions due to the relatively weak quiet-Sun magnetic fluxes. The mass and magnetic free energy can be transferred from the lower corona to the middle or higher corona, which could become a possible source of the solar wind. \citet{2023Chitta} gave a direct observational evidence for a coronal separatrix web driving highly structured slow solar wind through complex middle-coronal dynamic processes. Perhaps the minifilament eruption we observed could transport mass and energy to the middle or higher corona, which then becomes the solar wind through the mechanism proposed by their research.

\begin{deluxetable*}{llc|llc}
  \tablenum{1}
  \tablecaption{List of the small-scale eruptions with coronal dimmings from EUI/HRI$_{EUV}$ observations at 174 \AA}
  \tablewidth{0pt}
  \tablehead{
    \colhead{Event} & \colhead{Date/Time\tablenotemark{a}}  & \colhead{Location\tablenotemark{b} (arcsec)} & \colhead{Event} & \colhead{Date/Time}  & \colhead{Location (arcsec)}}
  \startdata
  1  & 2020-05-20/21:20   & -228, 288 & 10    & 2021-03-22/11:40  & 98, 556    \\
  2  & 2020-05-21/16:50	& -307, -203	& 11	& 2021-08-21/21:33	& -400, -5   \\
  3  & 2020-10-19/20:14	& 192, 147	& 12	& 2021-08-21/22:35	& 152, -68 \\
  4  & 2020-10-19/20:14	& 562, 71  & 13	& 2021-08-31/08:54	& 322, 172  \\
  5  & 2020-10-21/11:39	& 22, 209	& 14	& 2021-08-31/10:47	& 46, 204 \\
  6  & 2020-11-19/12:39	& -526, 375 & 15    & 2022-02-08/13:48	& -32, -115  \\
  7  & 2021-02-22/03:51	& -63, 117  & 16    & 2022-02-08/14:00	& -105, -85 \\
  8  & 2021-03-22/10:16	& -300, 313	& 17    & 2022-03-27/21:30	& -318, -123  \\
  9  & 2021-03-22/11:13	& -154, -119	\\
  \enddata
  \tablenotetext{a}{Approximate onset time of the eruptions.}
  \tablenotetext{b}{Helioprojective coordinates from the Solar Orbiter perspective.}
  \end{deluxetable*}

We thank Xudong Sun for providing suggestions on HMI data processing and D. Berghmans for valuable suggestions on EUI data. The research was supported by National Key R\&D Program of China No. 2021YFA0718600 and No. 2022YFF0503800, NSFC under grants 12073032, 42274201, 42004145 and 42150105, the Specialized Research Fund for State Key Laboratories of China. We acknowledge the use of data from Solar Orbiter and SDO. Solar Orbiter is a space mission of international collaboration between ESA and NASA, operated by ESA. The EUI instrument was built by CSL, IAS, MPS, MSSL/UCL, PMOD/WRC, ROB, LCF/IO with funding from the Belgian Federal Science Policy Office (BELSPO/PRODEX PEA 4000134088); the Centre National d'Etudes Spatiales (CNES); the UK Space Agency (UKSA); the Bundesministerium f\"{u}r Wirtschaft und Energie (BMWi) through the Deutsches Zentrum f\"{u}r Luft- und Raumfahrt (DLR); and the Swiss Space Office (SSO).


\appendix

\section{Mass-loss estimate of the eruption}

We use the DEM techniques to determine the CME masses from coronal dimmings. Considering the limitations posed by the signal-to-noise ratio of the AIA passbands, we only utilize the passbands centered at 171, 193, and 211 \AA, which are sufficient to encompass the temperature range of the quiet Sun within the heart-shaped area.

To determine the total evacuated mass $M$ of the eruption, we use
\begin{equation}
    \begin{split}
    M &= \sum_{i=0}^n \mu m_p A_s \int_0^L N(z)dz \\
      &= \sum_{i=0}^n \mu m_p A_s \lambda_p N_0 I\left(\frac{\lambda_p}{R_\odot}\right),
    \end{split}
\end{equation}
where $n$ is the total number of the pixels within the heart-shaped area, $\mu$ the mean molecular weight of the hydrogen ion ($\mu$ = 1.27), m$_p$ the proton mass, A$_s$ the area of each pixel, $\int_0^L N(z)dz$ the total number of coronal plasma at each pixel in the dimming region along the LOS, and $\lambda_{p}$ the pressure scale height defined as
\begin{equation}
    \lambda_p=\frac{2k_BT_e}{\mu m_pg_\odot},
\end{equation}
where $k_B$ is the Boltzmann constant, $T_e$ the electron temperature at each pixel calculated from the DEM analysis by Equation~\ref{tem}, and $g_\odot$ the gravitational acceleration at the solar surface.

We obtain the value of the basal electron density $N_{0}$ with the help of the EM measurement \citep[for detailed derivation, please refer to the previous implementation about CME mass determination of][]{2017Lopez}, i.e.,
\begin{equation}
    N_{0} = \sqrt{\frac{EM}{\frac{\lambda_{p}}{2} I\left(\frac{\lambda_{p}}{2R_\odot}\right)}},
\end{equation}
where $R_\odot$ is the solar radius, and $I$ the integral quantity defined as
\begin{equation}
    I(\alpha) = \int_0^{2L/\lambda_p} exp\left(-\frac{x}{\alpha x+1}\right)dx.
\end{equation}
We have set $L = 5\lambda_p$ in order to include as much mass as possible along the LOS.

The total evacuated mass $\Delta M$ is determined by the difference between the mass $M_{before}$ before the eruption and the mass $M_{after}$ after the eruption, i.e.,
\begin{equation}
   \Delta M = M_{before}-M_{after}.
\end{equation}




\end{CJK*}
\end{document}